\documentclass{article}

\usepackage{arxiv}
\usepackage[utf8]{inputenc} % allow utf-8 input
\usepackage[T1]{fontenc}    % use 8-bit T1 fonts
\usepackage{booktabs}       % professional-quality tables
\usepackage{amsfonts}       % blackboard math symbols
\usepackage{nicefrac}       % compact symbols for 1/2, etc.
\usepackage{microtype}      % microtypography
\usepackage{lipsum}		% Can be removed after putting your text content
\usepackage{amsmath,amssymb}
\usepackage{graphicx}
\usepackage{color}

\newcommand{\grl}{    {Geophys. Res. Lett.}}

\newcommand{\ssr}{    {Space Sci. Rev.}}

\newcommand{\aap}{    { Astronomy and Astrophysics}}

\newcommand{\apj}{ {Astrophys. J. }}
\newcommand{\apjl}{    {Astrophys. J. Lett.}}

\newcommand{\nat}{    {Nature}}

\newcommand{\prl}{    {Phys. Rev. Lett.}}
\newcommand{\apjs}{    {Astrophys. Journal. Suppl. Ser.}}

\def\XXint#1#2#3{{\setbox0=\hbox{$#1{#2#3}{\int}$}
     \vcenter{\hbox{$#2#3$}}\kern-.5\wd0}}

\title{Resonance of low-frequency electromagnetic and ion-sound modes in the solar wind}

\author{
	{I. Y. Vasko} \\
	William B. Hanson Center for Space Sciences, University of Texas at Dallas, Richardson, TX, USA\\
    \texttt{Ivan.Vasko@UTDallas.edu} \\\\
    {F. S. Mozer and T. Bowen} \\
	Space Sciences Laboratory, University of California at Berkeley, California, CA94720, USA\\\\
	{J. Verniero} \\
	NASA Goddard Space Flight Center, Washington, USA\\\\
  {X. An and A. V. Artemyev} \\
	Department of Earth, Planetary, and Space Sciences, University of California, Los Angeles, California, USA\\\\
  {J. W. Bonnell} \\
	Space Sciences Laboratory, University of California at Berkeley, Berkeley, California, USA\\\\
   {J. Halekas} \\
	Department of Physics and Astronomy, University of Iowa, Iowa, USA\\\\
    {I. V. Kuzihev} \\
	New Jersey Institute of Technology, Newark, New Jersey, USA
}

\begin{document}
\maketitle
\begin{abstract}
Parker Solar Probe measurements have recently shown that coherent fast magnetosonic and Alfv\'{e}n ion-cyclotron waves are abundant in the solar wind and can be accompanied by higher-frequency electrostatic fluctuations. In this letter we reveal the nonlinear process capable of channelling the energy of low-frequency electromagnetic to higher-frequency electrostatic fluctuations observed aboard Parker Solar Probe. We present Hall-MHD simulations demonstrating that low-frequency electromagnetic fluctuations can resonate with the ion-sound mode, which results in steepening of plasma density fluctuations, electrostatic spikes and harmonics in the electric field spectrum. The resonance can occur around the wavenumber determined by the ratio between local sound and Alfv\'{e}n speeds, but only in the case of {\it oblique} propagation to the background magnetic field.  The resonance wavenumber, its width and steepening time scale are estimated, and all indicate that the revealed two-wave resonance can frequently occur in the solar wind. This process can be a potential channel of energy transfer from cyclotron resonant ions producing the electromagnetic fluctuations to Landau resonant ions and electrons absorbing the energy of the higher-frequency electrostatic fluctuations.
\end{abstract}

% keywords can be removed

\section{Introduction}

Spacecraft measurements at 0.3--1 AU showed that circularly-polarized electromagnetic fluctuations with the frequency of the order of local proton cyclotron frequency can be continuously present in the solar wind for up to tens of minutes \cite{Jian09:apjl,Jian10:jgr,Jian14:apj,Boardsen15:jgr,Zhao18:jgr}. These low-frequency fluctuations are either left- or right-handed in the  spacecraft frame and propagate nearly parallel to local magnetic field lines. Refs. \cite{Jian10:jgr,Jian14:apj} interpreted these fluctuations in terms of Alfv\'{e}n ion-cyclotron waves, left-handed in the {\it plasma} frame. The observations of right-handed fluctuations in the spacecraft frame were suggested to be caused by the Doppler shift of Alfv\'{e}n ion-cyclotron waves propagating sunward in the plasma frame. The stability analysis of ion velocity distribution functions at 1 AU showed however that the low-frequency fluctuations can be also fast magnetosonic waves, right-handed in the {\it plasma} frame \cite{Gary16:jgr,Wicks16:apj}. The occurrence of the low-frequency fluctuations is probably much higher than a few percent reported at 0.3--1 AU \cite{Jian14:apj,Boardsen15:jgr,Zhao18:jgr}, since turbulence can mask their presence \cite{Gary&Podesta11,He11:apj}. The substantially larger occurrence would be consistent with the fact that ion velocity distribution functions in the solar wind are typically unstable to low-frequency electromagnetic fluctuations \cite{Klein18:prl}.

\begin{figure*}[ht!]
\centering
\includegraphics[scale=0.55,angle=0,origin=c]{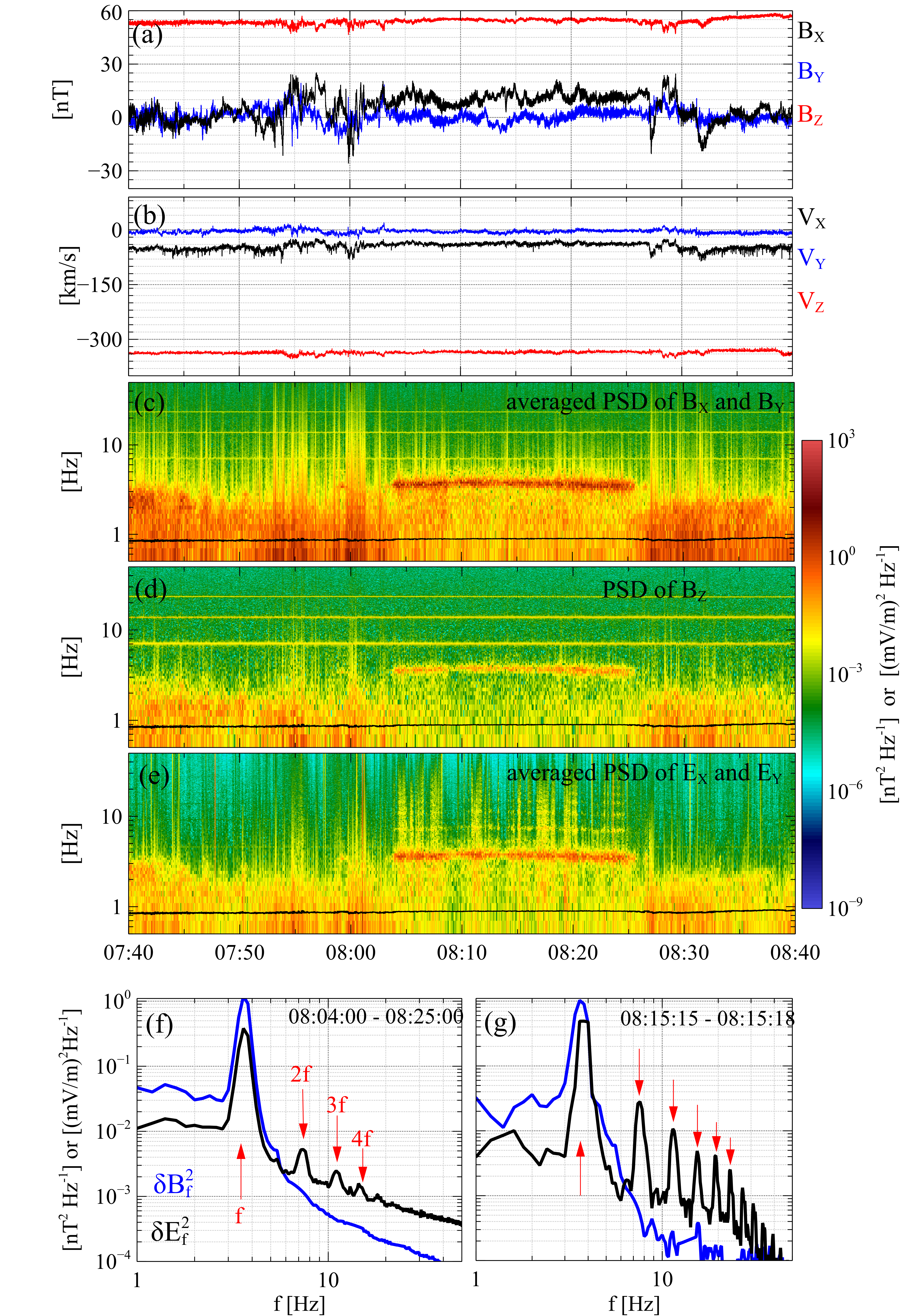}
\caption{Overview of PSP observations on April 1, 2019: (a) the magnetic field measured at about 146 S/s resolution and (b) the ion flow velocity provided at 0.2s cadence; (c) the mean of power spectral densities (PSD) of magnetic field components in the $XY$ plane and (d) the PSD of the magnetic field component along the $Z$-axis that is directed radially toward the Sun; (e) the mean of PSDs of electric field components in the $XY$ plane; (f, g) omnidirectional spectra of three magnetic and two electric field components averaged over about 20 minute and 3s intervals. Note that the instrumental noise at about 8, 16 and 32 Hz noticeable in panel (d) has been removed in panels (f) and (g). The PSDs were computed using the Fast Fourier Transform with a 5s sliding window and Hanning window function.} 
\label{fig1}
\end{figure*}

Parker Solar Probe (PSP) and Solar Orbiter measurements have recently shown that similar low-frequency fluctuations are abundant closer to the Sun \cite{Bale19:nature,Bowen20:apjs,Khotyaintsev21:aa}. The occurrence of these fluctuations is around 30\% within 0.3 AU \cite{Liu23:apj_radial_evolution}. The analysis of particle distribution functions showed that the low-frequency fluctuations can be both Alfv\'{e}n ion-cyclotron and fast magnetosonic waves produced by local ion-driven instabilities \cite{Verniero20,Klein21:apj}. In contrast to earlier analyses at 0.3--1 AU, PSP has provided electric field measurements that allowed quantifying Doppler shifts and establishing that the low-frequency fluctuations can be indeed left- and right-handed in the {\it plasma} frame \cite{Mozer20:jgr,Bowen20:apj}. The low-frequency fluctuations may contribute to solar wind heating \cite{Bowen22:prl,Ofman22:apj}, but specific dissipation mechanisms are still investigated. Using electric field measurements, Ref. \cite{Verniero20} have shown that the low-frequency electromagnetic fluctuations can be accompanied by electrostatic fluctuations at higher frequencies. They hypothesized that some nonlinear process channels the energy from the electromagnetic to electrostatic fluctuations, but the physics of that nonlinear process has not been revealed. 

%Parametric instabilities \cite{Matteini10:jgr} and local instabilities induced by the electromagnetic fluctuations \cite{Valentini14:apj,Roytershteyn21:phpl} are not among likely candidates. 

In this letter we demonstrate that electrostatic fluctuations reported by Ref. \cite{Verniero20} are produced due to resonance of slightly oblique electromagnetic (fast magnetosonic) fluctuations with the ion-sound plasma mode. The resonance results in steepening of low-frequency plasma density fluctuations, electrostatic spikes phase-correlated with the electromagnetic fluctuations and harmonics in the electric field spectrum. The resonance condition, its width and steepening time scale are estimated. The potential effect of this nonlinear process in the dissipation of low-frequency electromagnetic fluctuations is discussed.

\section{Data and observations\label{sec1}}

We use magnetic and electric field measurements provided at respectively about 146 and 293 S/s (Samples per second) by the FIELDS instrument suite \cite{Bale16:ssr}, ion and electron moments provided at respectively 0.2 and 28s cadence by the SWEAP instrument suite \cite{Kasper16:ssr}, and plasma density estimates at 7s cadence delivered by the quasi-thermal noise spectroscopy \cite{Moncuquet20:apj}. The electric field aboard PSP is measured by four 2-meter long cylindrical whip sensors mounted at the edges of the forward heat shield and providing two electric field components in the $XY$ plane perpendicular to the $Z$ axis that is the sunward radial direction. The electric field is calibrated through a least-squares fit to $-\textbf{\emph{V}}_{i}\times \textbf{\emph{B}}_0$, where ion flow velocity $\textbf{\emph{V}}_{i}$ and background magnetic field $\textbf{\emph{B}}_0$ are averaged over 12s \cite{Mozer20:jgr}. The effective antenna length at the frequencies of interest, 1--10 Hz, is between about 0.5 and 2 of its DC value revealed by the least squares fit \cite{Bowen20:apj}. 

Figure \ref{fig1} overviews a one hour interval of PSP observations on April 1, 2019. Panels (a) and (b) show that the magnetic field of about 55 nT was directed almost radially toward the Sun, while the plasma was flowing radially outward at about 340 km/s. The plasma density was around 230 cm$^{-3}$ during the entire interval according to the quasi-thermal noise spectroscopy (not shown). The electron and ion temperatures were both around 30 eV (not shown), though ion velocity distributions were not Maxwellian \cite{Verniero20}, the corresponding electron and ion betas were $\beta_{e}\approx \beta_{i}\approx 1$. Panels (c) and (d) present power spectral densities (PDS) of magnetic field components in the $XY$ plane and along the radial direction. The PSDs demonstrate that narrow-band magnetic field fluctuations around 4 Hz were continuously present for about 20 minutes. These fluctuations must have propagated within a few degrees of the background magnetic field, since the PSD of the radial component was a hundred times lower than that of the components in the $XY$ plane. Panel (e) presents the PSD of two electric field components and demonstrates the expected narrow-band electric field fluctuations around 4 Hz. These fluctuations are accompanied by the wave power at harmonics of the fundamental frequency of 4 Hz. Panel (f) shows distinct harmonics in the electric field PSD averaged over the 20 minute interval, but no similar harmonics in the averaged magnetic field PSD. There are only three distinct harmonics visible, but panel (g) demonstrates up to ten harmonics in a PSD computed over a 3s interval.

\begin{figure*}[ht!]
\centering
\includegraphics[scale=0.5,angle=0,origin=c]{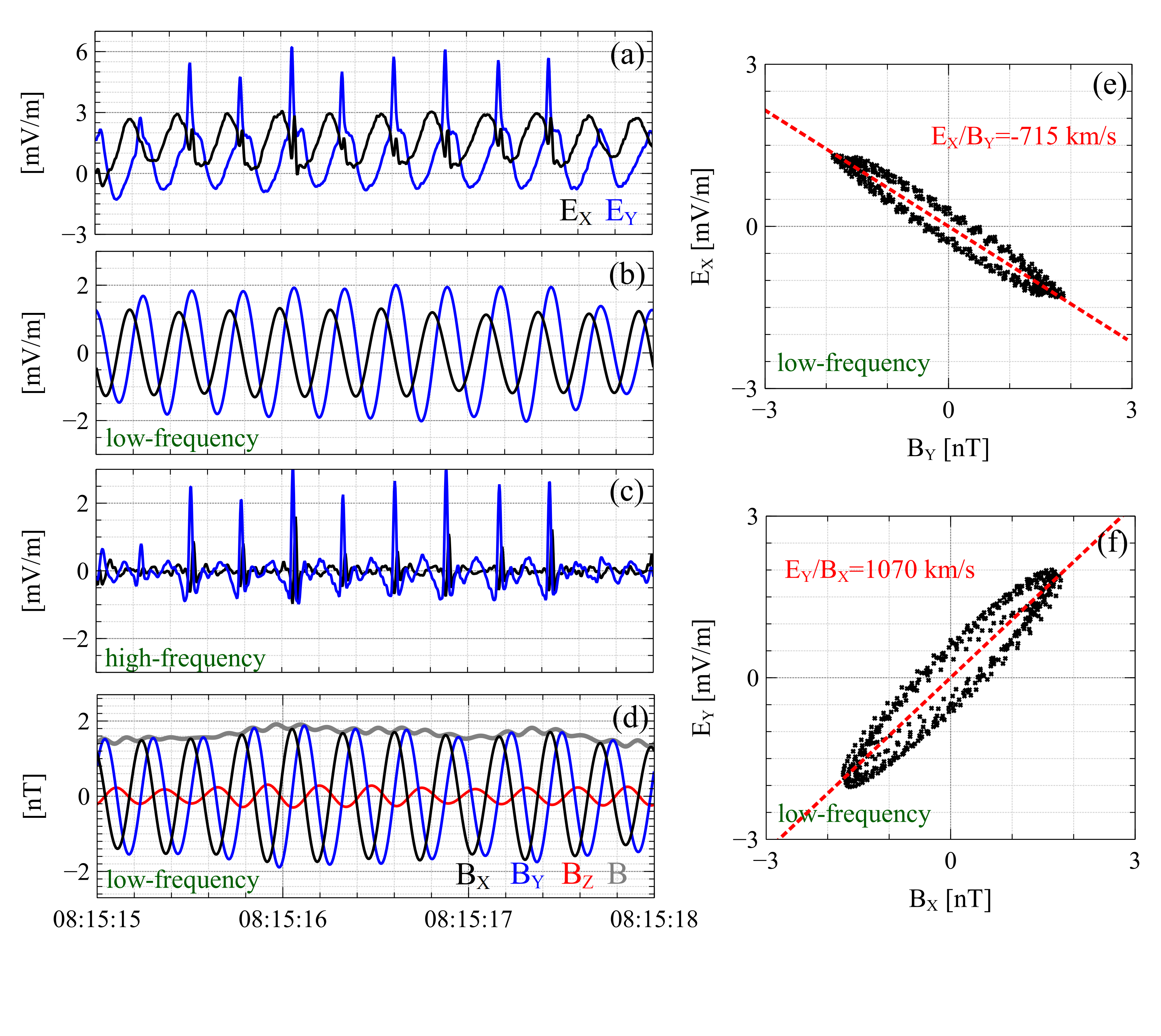}
\caption{An expanded view of electric and magnetic fields measured over a 3s interval: (a) waveforms of two electric field components measured at about 293 S/s resolution; (b) low-frequency and (c) high-frequency electric field fluctuations computed by band-pass filtering the electric field over 2--6 Hz and 6--146 Hz; (d) similarly computed low-frequency magnetic field. Panels (e) and (f) compare low-frequency electric and magnetic fields in the $XY$ plane, and indicate the best-fit slopes for $E_{X}$ vs. $B_{Y}$ and $E_{Y}$ vs. $B_{X}$.} 
\label{fig2}
\end{figure*}

\begin{figure*}[ht!]
\centering
\includegraphics[scale=0.5,angle=0,origin=c]{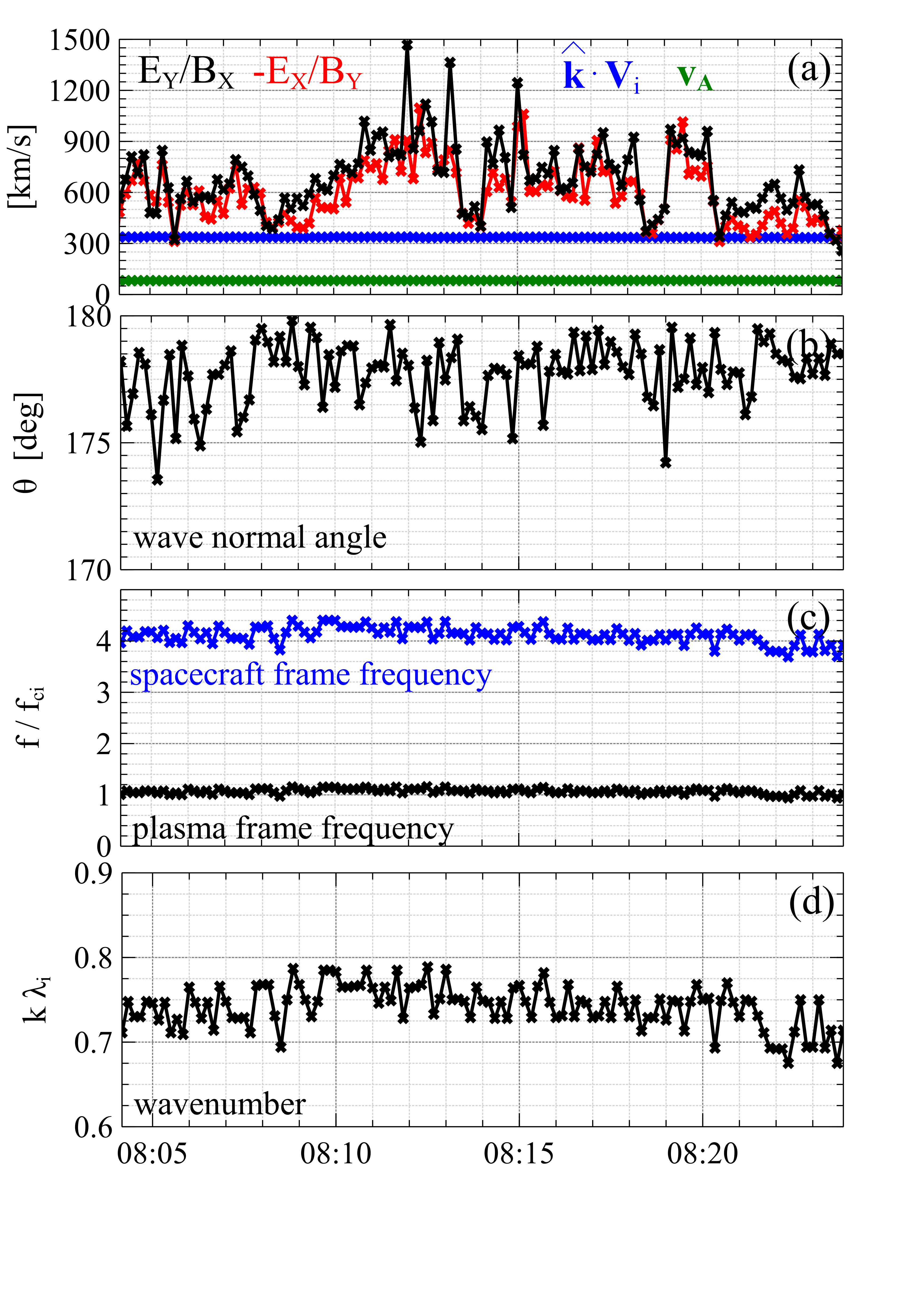}
\caption{The properties of low-frequency, band-pass filtered over 2--6 Hz,  electric and magnetic fields. For each 10s subinterval of about 20 minute interval we computed: (a) the best fit slopes for $E_{X}$ vs. $B_{Y}$ and $E_{Y}$ vs $B_{X}$, Alfv\'{e}n speed $v_{A}$ and $\hat{\textbf{\emph{k}}}\cdot \textbf{\emph{V}}_{i}$; the latter is the ion flow velocity component along unit vector $\hat{\textbf{\emph{k}}}$ that is the propagation direction of the low-frequency electromagnetic fluctuations; (b) the wave normal angle between $\hat{\textbf{\emph{k}}}$ and local background magnetic field; (c) plasma and spacecraft frame frequencies corresponding to the maximum power spectral density and (d) the corresponding wavenumber. The frequencies and wavenumbers are normalized to local ion cyclotron frequency $f_{ci}$ and ion inertial length $\lambda_{i}$, respectively.} 
\label{fig3}
\end{figure*}

Figure \ref{fig2} presents electric and magnetic field waveforms observed over that 3s interval. Panel (a) demonstrates electric field waveforms, while panels (b) and (c) show its low- and high-frequency constituents computed by band-pass filtering over 2--6 Hz and 6--146 Hz, respectively. Panel (d) presents similarly computed low-frequency magnetic field. The fluctuations around 4 Hz are circularly-polarized and right-handed, both electric and magnetic fields rotate in the direction of electron gyration. In contrast, the high-frequency fluctuations consist of linearly-polarized electric field spikes phase-correlated with the low-frequency electromagnetic fluctuations. It is these {\it electrostatic} spikes that create harmonics in the electric field spectra. Panels (e) and (f) demonstrate that the low-frequency electric and magnetic fields in the $XY$ plane are correlated and the best-fit slopes are $E_{X}/B_{Y}\approx-715$ km/s and $E_{Y}/B_{X}\approx 1070$ km/s. Since the low-frequency fluctuations propagate almost perpendicular to the $XY$ plane, these correlations imply {\it outward} propagation from the Sun in the spacecraft frame. The {\it actual} phase speed estimate is however dependent on the effective antenna length to be used.

Figure \ref{fig3} presents properties of the low-frequency electromagnetic fluctuations. For each 10s subinterval of about 20 minute interval, we estimated $E_{X}/B_{Y}$ and $E_{Y}/B_{X}$, and also applied the Minimum Variance Analysis \cite{Sonnerup&Scheible98} to compute unit vector $\hat{\textbf{\emph{k}}}$ along the minimum variance direction that is equivalent to the propagation direction. Panels (a) and (b) show that $-E_{X}/B_{Y}$ and $E_{Y}/B_{X}$ are consistent and typically within 400--900 km/s, while the propagation direction is within 175$^{\circ}$--180$^{\circ}$ of the background magnetic field. The electromagnetic fluctuations are either fast-magnetosonic waves propagating radially outward or Alfv\'{e}n ion-cyclotron waves propagating sunward in the {\it plasma} frame. The phase speed of slightly oblique fast magnetosonic and Alfv\'{e}n ion-cyclotron waves in the {\it plasma} frame is respectively larger and smaller than the Alfv\'{e}n speed $v_{A}$ (Section \ref{sec2}). Therefore, the phase speed in the spacecraft frame would exceed $\textbf{\emph{V}}_{i}\cdot \hat{\textbf{\emph{k}}}+v_{A}\approx 420$ km/s in the case of fast magnetosonic waves, while would be smaller than $\textbf{\emph{V}}_{i}\cdot \hat{\textbf{\emph{k}}}-v_{A}\approx 260$ km/s in the case of Alfv\'{e}n ion-cyclotron waves, where $v_{A}\approx 80$ km/s and $\textbf{\emph{V}}_{i}\cdot \hat{\textbf{\emph{k}}} \approx 340$ km/s. The observed values of $E_{Y}/B_{X}\approx -E_{X}/B_{Y}\approx 400$--900 km/s indicate that the interpretation in terms of fast magnetosonic waves is consistent with the effective antenna length varying within a factor of two of its DC value. This interpretation is also consistent with linear instability of fast magnetosonic waves for several ion velocity distribution functions in the considered interval \cite{Verniero20}. Panels (c) and (d) present wavenumber $k\lambda_{i}$, spacecraft and plasma frame frequencies, $f_{sc}/f_{ci}$ and $f/f_{ci}$, where $f_{ci}$ and $\lambda_{i}$ are local ion cyclotron frequency and inertial length. The parameter $f_{sc}/f_{ci}\approx 4$ was determined by the peak value of the PSD, while $f/f_{ci}$ and $k\lambda_{i}$ were related through the cold plasma dispersion relation (Section \ref{sec2}) and computed using the Doppler-shift relation, $f_{sc}/f_{ci}=f/f_{ci}+k\lambda_{i}\textbf{\emph{V}}_{i}\cdot\hat{\textbf{\emph{k}}}/v_{A}$. Resolving this relation we found plasma frame frequency $f/f_{ci}\approx 1.2$ and wavenumber $k\lambda_{i}\approx 0.75$. In the next section we reveal the mechanism resulting in the formation of electrostatic spikes associated with fast magnetosonic waves. The schematics of the wave propagation direction in the coordinate system $XYZ$ is shown in Figure \ref{fig4}.

%in terms of Alfv\'{e}n ion-cyclotron waves would require the effective antenna length of two to four times larger than its DC value, which would be inconsistent with previous analyses . We interpret the low-frequency fluctuations in terms .

\section{Theoretical interpretation and numerical simulations\label{sec2}}

The electromagnetic fluctuations and electrostatic spikes observed aboard PSP have frequencies below ion plasma frequency and spatial scales above electron inertial length. In this case the plasma dynamics can be described within Hall-MHD approximation, which neglects electron inertia and assumes identical electron and ion densities (e.g., Ref. \cite{Formisano69}). The Hall-MHD equations can be written as follows
\begin{eqnarray} 
\frac{\partial N}{\partial t}&+&\nabla (N\textbf{\emph{V}})=0,\nonumber\\\nonumber\\
  \frac{\partial \textbf{\emph{V}}}{\partial t}+(\textbf{\emph{V}}\nabla)\textbf{\emph{V}}&=&-\frac{\nabla P}{m_{i}N}+\frac{{\rm curl}\;\textbf{\emph{B}}\times (\textbf{\emph{B}}_0+\textbf{\emph{B}})}{\mu_0m_{i}N},\nonumber\\ \\
  \frac{\partial \textbf{\emph{B}}}{\partial t}&=&-{\rm curl}\;\textbf{\emph{E}},\nonumber\\\nonumber\\\textbf{\emph{E}}+\textbf{\emph{V}}\times (\textbf{\emph{B}}_0+\textbf{\emph{B}})&=&-\frac{\nabla P_{e}}{eN}+\frac{{\rm curl}\;\textbf{\emph{B}}\times (\textbf{\emph{B}}_0+\textbf{\emph{B}})}{\mu_0 eN},\nonumber
  \label{eq:Hall_MHD}
\end{eqnarray}
where $N$ and $\textbf{\emph{V}}$ are the plasma density and flow velocity, $P=P_{e}+P_{i}$ is the total of electron and ion thermal pressures assumed to be isotropic, $\textbf{\emph{E}}$ and $\textbf{\emph{B}}_0+\textbf{\emph{B}}$ are electric and magnetic fields, $\textbf{\emph{B}}_0$ is a uniform background magnetic field. These equations are complemented by equations of state, $P_{e}/N^{\gamma_{e}}={\rm const}$ and $P_{i}/N^{\gamma_{i}}={\rm const}$, where $\gamma_{e}$ and $\gamma_{i}$ are polytrope indexes. We will demonstrate that it is the electron pressure gradient term in the Ohm’s law that results in the electrostatic spikes (Section \ref{sec1}).

Ref. \cite{Formisano69} presented the linear theory of waves in warm uniform plasma using the Hall-MHD approximation. In the plasma rest frame the corresponding linear dispersion relation between angular frequency $\omega$ and wavenumber $k$ of monochromatic waves propagating at wave normal angle $\theta$ to the background magnetic field reads
\begin{eqnarray}
  \left(\cos^{2}\theta-\frac{\omega^{2}}{k^{2}v_{A}^2}\right)\cdot\left(\cos^2\theta-\frac{\omega^{2}}{k^2c_{s}^{2}}-\frac{\omega^2}{k^2v_{A}^{2}}\left(1-\frac{\omega^{2}}{k^2c_{s}^{2}}\right)\right)=\frac{\omega^2}{\omega_{ci}^{2}}\cos^{2}\theta\left(1-\frac{\omega^{2}}{k^2c_{s}^2}\right),
  \label{eq:disp_rel}
\end{eqnarray}
where $\omega_{ci}=2\pi f_{ci}=eB_0/m_{i}$ is the angular ion cyclotron frequency. The sound speed $c_{s}$, Alfv\'{e}n speed $v_{A}$ and their ratio are given as follows
\begin{eqnarray}
  c_{s}=\left(\frac{\gamma_{e}T_{e}+\gamma_{i}T_{i}}{m_{i}}\right)^{1/2},\;\;\;\;\;\;v_{A}=\left(\frac{B_0}{\mu_0 N_0 m_{i}}\right)^{1/2},\;\;\;\;\;\;     \frac{c_{s}}{v_{A}}=\left(\frac{\gamma_{e}\beta_{e}+\gamma_{i}\beta_{i}}{2}\right)^{1/2},
  \label{eq:cs_and_va}
\end{eqnarray}
where $N_0$, $T_{e}$ and $T_{i}$ are background plasma density, electron and ion temperatures. In cold plasma, $\omega/kc_{s}\rightarrow \infty$, the dispersion relation simplifies to $(1-\omega^2/k^2v_{A}^2)(\cos^2\theta-\omega^2/k^2v_{A}^2)=\omega^2\cos^2\theta/\omega_{ci}^2$ and describes fast magnetosonic and Alfv\'{e}n ion-cyclotron modes. In warm plasma the dispersion relation already delivers three wave modes. In the case of strictly parallel propagation these are a purely electrostatic ion-sound mode, whose dispersion relation is $\omega=kc_{s}$, and purely electromagnetic fast magnetosonic and Alfv\'{e}n ion-cyclotron modes. In the case of oblique propagation the three modes are named fast, intermediate and slow, whose phase speeds $v_{F}$, $v_{I}$ and $v_{S}$ at $\omega/\omega_{ci}\rightarrow 0$ are ordered as $v_{S}<v_{I}<v_{F}$ and delivered by the classical expressions, $v_{I}=v_{A}\cos\theta$ and $v_{F,S}^{2}=(v_{A}^2+c_{s}^2)/2\pm\left((v_{A}^2+c_{s}^2)^2/4-v_{A}^2c_{s}^2\cos^2\theta\right)^{1/2}$. The asymptotic analysis of Eq. (\ref{eq:disp_rel}) at small wave normal angles, $\sin^{2}\theta\ll 1$, and low frequencies, $\sin^2\theta \ll \omega/\omega_{ci}\ll 1$, shows that in the case of $c_{s}>v_{A}$ the fast mode is practically dispersionless, while the intermediate and slow modes are dispersive
\begin{eqnarray}
 \left(\frac{\omega^2}{k^2}\right)_{F}\approx v_F^2,\;\;\;\;\;  \left(\frac{\omega^2}{k^2}\right)_{I}\approx v_{I}^2\left(1+\frac{\omega}{\omega_{ci}\cos\theta}\right),\;\;\;\;\;\left(\frac{\omega^2}{k^2}\right)_{S}\approx v_{S}^2\left(1-\frac{\omega\cos\theta}{\omega_{ci}}\right).
    \label{eq:dispersion}
\end{eqnarray}
Note that at small wave normal angles we have $v_{F}\approx c_{s}\left(1+v_{A}^2\sin^2\theta/2(c_{s}^2-v_{A}^2)\right)\approx c_{s}$, $v_{I}=v_{A}\cos\theta\approx v_{A}$, and $v_{S}\approx v_{A}\left(1-c_{s}^2\sin^2\theta/2(c_{s}^2-v_{A}^2)\right)\approx v_{A}$.

\begin{figure*}[ht!]
\centering
\includegraphics[scale=0.25,angle=0,origin=c]{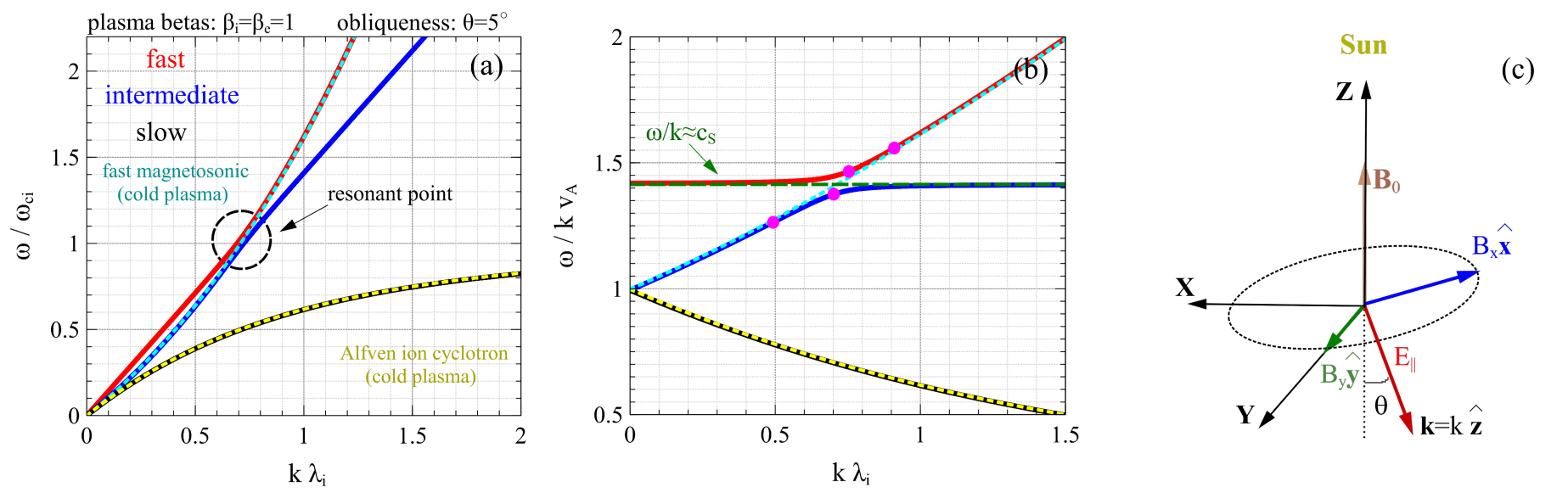}
\caption{Panels (a) and (b) present linear dispersion curves of fast, intermediate and slow modes obtained using Hall-MHD approximation for the case of oblique propagation at the wave normal angle of $\theta=5^{\circ}$ in a  warm plasma with $\beta_{e}=\beta_{i}=1$, and polytrope indexes of electrons and ions of $\gamma_{e}=1$ and $\gamma_{i}=3$. Both panels also demonstrate dispersion curves of fast magnetosonic and Alfv\'{e}n ion-cyclotron modes in cold plasma. Panel (c) presents the coordinate system $XYZ$ used for PSP observations in Section \ref{fig1}: the $Z$ axis is directed radially toward the Sun and basically parallel to the background magnetic field (Figure \ref{fig1}). Panel (c) also presents the coordinate system $xyz$ used for numerical simulations in Section \ref{sec3}: the $z$-axis is along the propagation direction $\hat{\textbf{\emph{k}}}$ of the low-frequency electromagnetic fluctuations. The magenta dots in panel (b) correspond to fast magnetosonic waves, whose nonlinear evolution is demonstrated in Figure \ref{fig5}.} 
\label{fig4}
\end{figure*}

%Whatever reasonable values of polytropic indexes $\gamma_{e}$ and $\gamma_{i}$ are used, in a warm plasma with $\beta_{e}\approx@we would always obtain $c_{s}/v_{A}\gtrsim 1$  
%reasonable values of $\gamma_{e}$ and $\gamma_{i}$ we assume, once electron and ion betas are around one we have $c_{s}\gtrsim v_{A}$.

Figure \ref{fig4} presents the dispersion curves of fast, intermediate and slow modes propagating at wave normal angle of $\theta=5^{\circ}$ in a warm plasma with $c_{s}/v_{A}\approx 1.4$, which corresponds to $\beta_{e}=\beta_{i}=1$ and polytrope indexes of $\gamma_{e}=1$ and $\gamma_{i}=3$. Panel (a) shows $\omega/\omega_{ci}$ versus $k\lambda_{i}$, while panel (b) demonstrates $\omega/kv_{A}$ versus $k\lambda_{i}$, where $\lambda_{i}$ is the ion inertial length. Both panels also present the dispersion curves of fast magnetosonic and Alfv\'{e}n ion-cyclotron waves in cold plasma. The slow mode can be clearly associated with the Alfv\'{e}n ion-cyclotron wave, while neither fast nor intermediate modes alone can be associated with the fast magnetosonic wave. The dispersion curve of the intermediate mode is almost identical with that of the fast magnetosonic wave at $k\lesssim k_{*}$, while becomes close to $\omega=kc_{s}$ at $k\gtrsim k_{*}$. In turn, the dispersion curve of the fast mode is close to $\omega=kc_{s}$ at $k\lesssim k_{*}$ and almost identical to that of the fast magnetosonic wave at $k\gtrsim k_{*}$. The intermediate and fast modes are closely approaching each other at $k\approx k_{*}$ that can be named a resonance point between the ion-sound and fast magnetosonic waves. Assuming $(\omega^2/k^2)_{F}\approx (\omega^2/k^2)_{I}$ and using Eq. (\ref{eq:dispersion}) we estimate the wavenumber and frequency of the resonance point 
\begin{eqnarray}
    \frac{\omega_{*}}{\omega_{ci}}\approx \cos\theta \left(\frac{v_{F}^2}{v_{I}^2}-1\right)\approx \frac{c_{s}^2}{v_{A}^2}-1,\;\;\;\;\;\;\;k_{*}\lambda_{i}\approx \frac{v_{F}}{v_{I}}-\frac{v_{I}}{v_{F}}\approx \frac{c_{s}}{v_{A}}-\frac{v_{A}}{c_{s}},
    \label{eq:coupling_point}
\end{eqnarray}
where we have taken into account that $\sin^2\theta \ll 1$. In the considered case of $c_{s}/v_{A}\approx 1.4$ we obtain $\omega_{*}\approx \omega_{ci}$ and $k_{*}\lambda_{i}\approx 0.7$ that is consistent with panels (a) and (b).

Below we use coordinate system \textbf{\emph{xyz}} shown in panel (c) with unit vector \textbf{\emph{z}} along the propagation direction $\hat{\textbf{\emph{k}}}$ and unit vector \textbf{\emph{x}} in the plane of $\hat{\textbf{\emph{k}}}$ and $\textbf{\emph{B}}_0$. Note that unit vector $\hat{\textbf{\emph{k}}}$ is oriented at angle $\theta$ to $-\textbf{\emph{B}}_0$. The electric and magnetic field fluctuations $\textbf{\emph {E}}$ and $\textbf{\emph{B}}$ in Eqs. (\ref{eq:Hall_MHD}) can be decomposed into electromagnetic components $\textbf{\emph {E}}_{\perp}$ and $\textbf{\emph{B}}_{\perp}$ in the \textbf{\emph{xy}} plane, and the electrostatic component $E_{||}$ along the wave vector. Note that $B_{||}=0$, because ${\rm div}\textbf{\emph{B}}=0$. The  plasma flow velocity is similarly decomposed into compressible and incompressible components, $V_{||}$ and $\textbf{\emph{V}}_{\perp}$. The analysis of linear monochromatic fluctuations, whose quantities are proportional 
to $e^{i(kz-\omega t)}$, reveals the following relations
\begin{eqnarray}
&&\frac{N}{N_0}=1+\frac{kv_{A}}{\omega}\frac{V_{||}}{v_{A}},\;\;\;\;\frac{V_{||}}{v_{A}}=-\frac{kv_{A}\sin\theta/\omega}{1-k^2c_{s}^2/\omega^2}\frac{B_{x}}{B_0},\label{eq:Vpar}\\\nonumber\\
&&\frac{\textbf{\emph{V}}_{\perp}}{v_{A}}=-\frac{k v_{A}\cos\theta}{\omega}\frac{\textbf{\emph{B}}_{\perp}}{B_0},\;\;\;\frac{\textbf{\emph{E}}_{\perp}}{v_{A}B_0}=-\frac{\omega}{kv_{A}} \textbf{\emph{z}}\times \frac{\textbf{\emph{B}}_{\perp}}{B_0},\label{eq:Vper}\\\nonumber\\
&&\frac{B_{y}}{B_0}=\frac{i\omega\cos\theta/\omega_{ci}}{\omega^2/k^2v_{A}^2-\cos^2\theta}\frac{B_{x}}{B_0},\;\;\;\;\label{eq:pol_B}\\\nonumber\\
    \;\;\;\;\;\;\;\;\;\;
&&\frac{E_{||}}{v_{A}B_0}=\tan\theta\left(
    \frac{\omega}{kv_{A}}+\frac{\gamma_{e}\beta_{e}}{2}\frac{kv_{A}}{\omega}\frac{\omega^2-k^2v_{A}^2\cos^2\theta}{\omega^2-k^2c_{s}^2}\right)\frac{B_{y}}{B_0}\label{eq:Epar}.
\end{eqnarray}
We consider the intermediate mode at $k\lesssim k_{*}$ and the fast mode at $k\gtrsim k_{*}$, whose combination will be referred as the fast magnetosonic wave. Eq. (\ref{eq:pol_B}) shows that the fast magnetosonic wave is right-handed $B_{y}\propto iB_{x}$, because $\omega^2> k^2v_{A}^2\cos^2\theta$, and should be almost circularly polarized at $\omega/\omega_{ci}\gg \sin^2\theta$ as in cold plasma, because $B_{y}/B_{x}$ is independent of the sound speed. The sound speed enters into electrostatic component $E_{||}$ through the electron pressure gradient term in the Ohm's law and into the compressible velocity component, making the ratio $V_{||}/V_{\perp}$ and plasma density fluctuations by a factor of $(1-k^2c_{s}^2/\omega^2)^{-1}$ larger compared to cold plasma. The electrostatic component and compressibility are therefore expected to enhance at $k\approx k_{*}$.

%, where the phase speed of the fast magnetosonic wave is close to the sound speed. 

The nonlinear evolution of 1D planar fast magnetosonic waves propagating oblique to the background magnetic field is addressed by numerically solving Eqs. (\ref{eq:Hall_MHD}) with $(\textbf{\emph{V}}\nabla)\rightarrow V_{||}\; \partial / \partial z$, $(\textbf{\emph{B}}_0\nabla)\rightarrow B_0\cos\theta\;\partial/\partial z$ and $(\textbf{\emph{B}}\nabla)\rightarrow 0$. The initial condition is a monochromatic fast magnetosonic wave with $B_{x}|_{t=0}=\delta B \cos(k z)$ and other quantities delivered by Eqs. (\ref{eq:Vpar})--(\ref{eq:Epar}) of the linear theory. All the computations are carried out in the reference frame of the fast magnetosonic wave that is accomplished by replacing $\partial /\partial t\rightarrow \partial /\partial t-\omega/k$. The Hall-MHD equations describing evolution of plasma density $N$, flow velocity $\textbf{\emph{V}}$ and magnetic field $\textbf{\emph{B}}_{\perp}$  are solved in the spatial domain of $0\leq z\leq \lambda=2\pi/k$ with periodic boundary conditions. The numerical scheme based on the pseudo-spectral method that was previously applied for similar equations \cite{Vasko18:prl,Dillard18:pop} conserves the total energy up to 0.01\%. The numerical simulations were carried out for fast magnetosonic waves with initial amplitude $\delta B=0.04B_0$, propagating at wave normal angle of $\theta=5^{\circ}$ in a warm plasma with $\beta_{e}=\beta_{i}=1$ and polytrope indexes of $\gamma_{e}=1$ and $\gamma_{i}=3$. The initial wavenumber $k\lambda_i$ was chosen close to the coupling point of $k_{*}\lambda_i\approx 0.7$ (Figure \ref{fig4}).

\begin{figure*}[ht!]
\centering
\includegraphics[scale=0.05,angle=0,origin=c]{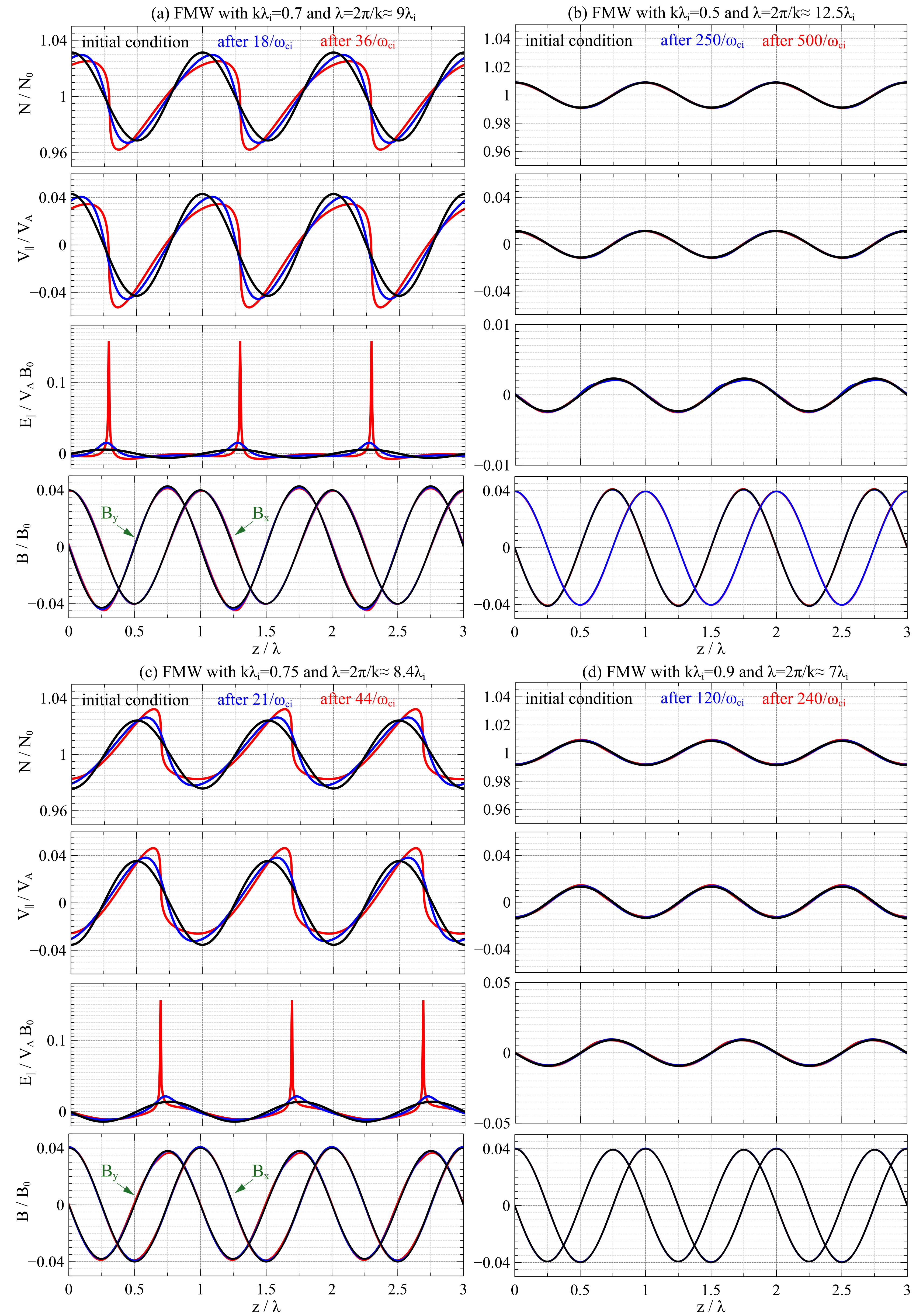}
\caption{The results of Hall-MHD simulations of the nonlinear evolution of initially monochromatic fast magnetosonic waves, propagating at $\theta=5^{\circ}$ to the background magnetic field in a warm plasma with $\beta_{e}=\beta_{i}= 1$. The initial wavenumbers of the fast magnetosonic waves are indicated in the panels as well as in Figure \ref{fig4}. For each simulation run the panels demonstrate evolution (in the frame of the wave) of the plasma density $N/N_0$, compressible component $V_{||}/v_{A}$ of the plasma flow velocity, electrostatic field $E_{||}/v_{A}B_0$ parallel to the propagation direction, and the magnetic field $\textbf{\emph{B}}/B_0$.} 
\label{fig5}
\end{figure*}

Figure \ref{fig5} demonstrates the nonlinear evolution of plasma density $N/N_0$, compressible velocity component $V_{||}/V_{A}$, electrostatic field $E_{||}/v_{A}B_0$ and magnetic field $\textbf{\emph{B}}_{\perp}/B_0$. The simulation results are presented in the domain encompassing three wavelengths to stress periodicity. Panels (a) show the evolution of the intermediate mode with $k\lambda_{i}=0.7$, while panels (b) demonstrate similar evolution of the fast mode with $k\lambda_{i}=0.75$. In both cases the waves are right-handed and circularly-polarized, the initial amplitude of the electrostatic component $E_{||}/v_{A}B_0\approx 5\cdot 10^{-3}$ is one order of magnitude smaller than the amplitude of the electromagnetic component, $E_{\perp}/v_{A}B_0=(\omega/kv_{A})\;B_{\perp}/B_0\approx 0.05$. The compressible component of the plasma flow velocity is initially comparable with the incompressible component, $V_{||}/v_{A}\approx 0.04$ and $V_{\perp}/v_{A}\approx (kv_{A}\cos\theta/\omega) B_{\perp}/B_0 \approx 0.03$, and the amplitude of density fluctuations is around $0.04N_0$. Both these quantities are about ten times larger than would be in cold plasma. Panels (a) and (b) show that the magnetic field $\textbf{\emph{B}}_{\perp}$ is essentially not evolving with time and the same is true for electromagnetic components $\textbf{\emph{E}}_{\perp}$ and incompressible velocity component $\textbf{\emph{V}}_{\perp}$ (not shown). In quite a contrast, a substantial evolution has occurred by the time of about $40\;\omega_{ci}^{-1}$ in the plasma density, compressible velocity component and electrostatic field. The observed evolution of $N/N_0$ and $V_{||}/V_{A}$ resembles steepening of sound waves in fluids and plasmas (e.g., Ref. \cite{Gurbatov83:Uspekhi}). The steepened plasma density fluctuations result in spikes in the electrostatic field $E_{||}$ due to the electron pressure gradient term $\nabla P_{e}/eN$ in the Ohm's law. Numerical computations well beyond $40\;\omega_{ci}^{-1}$ led to development of gradient scales smaller than the grid size of $\lambda/300$. The use of a smaller grid size allowed extending simulations only slightly beyond $40\;\omega_{ci}^{-1}$ and resulted in development of even smaller gradient scales and larger-amplitude spikes (not shown). This indicates of wave breaking within finite time (e.g. Ref. \cite{Gurbatov83:Uspekhi}).

The critical question is whether steepening would occur for initial wavenumbers sufficiently far from the resonance point. Panels (c) and (d) present the evolution of the intermediate and fast modes with respectively $k\lambda_{i}=0.5$ and 0.9. In both cases the amplitude of plasma density fluctuations is about 0.01$N_0$. The amplitude of the compressible velocity component $V_{||}/v_{A}\approx 0.01$ is smaller than the amplitude of the incompressible component, $V_{\perp}/v_{A}\approx 0.03$, but still larger than in cold plasma, where according to Eqs. (\ref{eq:Vpar}) and (\ref{eq:Vper}) we would have $V_{||}/V_{\perp}\approx \tan\theta\approx 0.1$. Both panels demonstrate that even by the time of $500\;\omega_{ci}^{-1}$ none of the quantities exhibit nonlinear evolution.  This strongly indicates that steepening and development of electrostatic spikes is a resonant phenomenon with a finite resonance width, $\Delta k=|k-k_{*}|$. Since additional simulations carried out for the intermediate and fast modes with respectively $k\lambda_{i}=0.6$ and 0.8 revealed steepening, while similar simulations for $k\lambda_{i}=0.55$ and 0.85 revealed no steepening (not shown), we can state that in the considered case we have $0.1\lesssim \Delta k\lambda_{i}\lesssim 0.15$.

The physics of the process revealed in Figure \ref{fig5} can be clarified by noticing that the nonlinear evolution occurs only for the plasma density and compressible velocity component. The evolution of these quantities can be described by two of the Hall-MHD equations
\begin{eqnarray}
    \frac{\partial N}{\partial t}+\frac{\partial \left(N V_{||}\right)}{\partial z}=0,\;\;\;\;\;\;m_{i}N\left[\frac{\partial V_{||}}{\partial t}+V_{||}\frac{\partial V_{||}}{\partial z}\right]=-\frac{\partial P}{\partial z}+F(\chi),
    \label{eq:reduced}
\end{eqnarray}
where $P=N_{0}T_{e} (N/N_0)^{\gamma_{e}}+N_{0}T_{i} (N/N_0)^{\gamma_{i}}$, while $F(\chi)={\rm curl}\;\textbf{\emph{B}}_{\perp}\times (\textbf{\emph{B}}_0+\textbf{\emph{B}}_{\perp})/\mu_0$ is determined by the initial magnetic field of the fast magnetosonic wave and depends only on $\chi=kz- \omega t$. Eqs. (\ref{eq:reduced}) without $F(\chi)$ would predict that any initially monochromatic wave propagates at the sound speed $c_{s}$ and breaks within a finite time of $\tau_{S}=\left[{\rm max}\;|\partial V_{||}/\partial z|_{t=0}\right]^{-1}=(k\delta V)^{-1}$, where $\delta V$ is the initial amplitude of $V_{||}$ (e.g., \cite{Gurbatov83:Uspekhi}). The simulations demonstrated that similar steepening occurs in the presence of $F(\chi)$ provided that it is resonant, $\omega/k\approx c_{s}$. The time scale of $40\;\omega_{ci}^{-1}$ revealed in Figure \ref{fig5} is quite close to $\tau_{S}=(k\delta V)^{-1}\approx 35\;\omega_{ci}^{-1}$, where we used $k\lambda_{i}=0.7$ and $\delta V/V_{A}=0.04$. The steepening is expected to vanish, when $F(\chi)$ is not resonant, $|\omega/k-c_{s}|\tau_{S}\gtrsim \lambda/2$ or $|\omega/k-c_{s}|\gtrsim \pi \delta V$, so that perturbations propagating at $\omega/k$ and $c_{s}$ cannot remain in phase over the time scale of $\tau_{S}$. Since according to Eq. (\ref{eq:Vpar}) we have $\delta V\propto \delta B\;(1-k^2c_{s}^2/\omega^2)^{-1}$, the resonance width $\Delta k$ resulting from $|\omega/k-c_{s}|\approx \pi \delta V$ is more appropriate to be expressed through the magnetic field amplitude 
\begin{eqnarray}
\left(\frac{\partial \omega}{\partial k}-\frac{\omega}{k}\right)_{k=k_*}^2\frac{\Delta k^2}{k_*^2}\approx \pi v_{A}^2\;\frac{\delta B\sin\theta}{B_0}\frac{\omega/k}{\omega/k+c_{s}},
\label{eq:dk_1}
\end{eqnarray}
where we took into account that $\omega/k-c_{s}= \omega/k-\omega_*/k_*\approx \left(\partial \omega/\partial k-\omega/k\right)_{k=k_*} \Delta k/k_*$. Restricting the analysis to $\sin^2\theta\ll 1$ we use the dispersion relation of strictly parallel fast magnetosonic waves, $\omega^2=k^2v_{A}^2 (1+\omega/\omega_{ci})$, to estimate the left-hand side of Eq. (\ref{eq:dk_1}). Substituting $\omega/k=c_{s}$ on the right-hand side of Eq. (\ref{eq:dk_1}) and using Eqs. (\ref{eq:coupling_point}) for $k_{*}\lambda_i$ and $\omega_{*}/\omega_{ci}$, we obtain the resonance width
\begin{eqnarray}
  \Delta k\lambda_{i}\approx \left(\frac{\pi}{2}\right)^{1/2}\left(1+\frac{v_{A}^2}{c_{s}^2}\right)\left(\frac{\delta B\sin\theta}{B_0}\right)^{1/2}.
  \label{eq:res_width}
\end{eqnarray}
For the case presented in Figure \ref{fig5}, where $\delta B/B_0=0.04$, $\theta=5^{\circ}$ and $c_s\approx 1.4 v_{A}$, Eq. (\ref{eq:res_width}) delivers the resonance width of $\Delta k\lambda_{i}\approx 0.11$ that is consistent with the simulation results. Similar calculations for the steepening time scale $\tau_{S}=(k\delta V)^{-1}$ lead to
\begin{eqnarray}
    \tau_{S}=\frac{\pi}{\omega_{ci}}\;\frac{(1+v_{A}^2/c_s^2)}{k_*\lambda_i\;\Delta k\lambda_i}\cdot \frac{|k-k_*|}{\Delta k}.
    \label{eq:steep_time}
\end{eqnarray}
Note that Eq. (\ref{eq:steep_time}) is valid only for $|k-k_*|\lesssim \Delta k$, while no steepening occurs for $|k-k_{*}|\gtrsim \Delta k$. For the case presented in Figure \ref{fig5}, Eq. (\ref{eq:steep_time}) predicts $\tau_{S}\approx 60\;\omega_{ci}^{-1}\cdot |k-k_{*}|/\Delta k$ that is consistent with the simulation results.

\section{Discussion and conclusion\label{sec3}}

The circularly-polarized electromagnetic fluctuations with the frequency of the order of local proton cyclotron frequency are abundant in the solar wind \cite{Boardsen15:jgr,Zhao18:jgr,Bowen20:apjs,Liu23:apj_radial_evolution}. These low-frequency electromagnetic fluctuations may contribute to solar wind heating \cite{Bowen22:prl,Ofman22:apj}, while corresponding dissipation mechanisms are still investigated. PSP measurements have recently shown that the low-frequency electromagnetic fluctuations can be accompanied by higher-frequency electrostatic fluctuations, which was assigned to some unknown nonlinear process potentially channeling the energy from lower to higher frequencies \cite{Verniero20}. Parametric instabilities \cite{Matteini10:jgr,Gonzalez20:apj,Gonzalez23:jpp} can be excluded, since no daughter waves were observed, while locally induced electrostatic instabilities \cite{Valentini11:apj,Valentini14:apj,Roytershteyn21:phpl} would not be consistent with highly coherent electric field waveforms.

We presented Hall-MHD simulations demonstrating that electrostatic fluctuations reported by \cite{Verniero20} can be produced due to the resonance of low-frequency (fast magnetosonic) fluctuations with the ion-sound mode. The resonance occurs only in the case of {\it oblique} propagation at wavenumbers $|k-k_{*}|\lesssim \Delta k$, where $k_{*}$ and $\Delta k$ are delivered by Eqs. (\ref{eq:coupling_point}) and (\ref{eq:res_width}). It manifests in the steepening of low-frequency plasma density fluctuations and results in electrostatic spikes due to the electron pressure gradient term in the Ohm's law (Eq. (\ref{eq:Hall_MHD})). In electric field spectra these electrostatic spikes, phase-correlated with the low-frequency electromagnetic fluctuations, correspond to wave power at harmonics of the fundamental frequency. No steepening or harmonics are observed in the magnetic field, because ${\rm curl} (\nabla P_{e}/N)=0$ and, hence, the electron pressure gradient term does not affect the magnetic field through $\partial \textbf{\emph{B}}/\partial t=-{\rm curl}\;\textbf{\emph{E}}$. The steepening time scale delivered by Eq. (\ref{eq:steep_time}) corresponds to the classical expression $\tau_{S}\approx (k\delta V)^{-1}$ (e.g., Ref. \cite{Gurbatov83:Uspekhi}), where $\delta V$ is the amplitude of low-frequency compressible velocity fluctuations. Note that for fast magnetosonic fluctuations the resonance can occur only when the sound speed is larger than the Alfv\'{e}n speed, $c_{s}>v_{A}$, but similar resonance can occur for Afv\'{e}n ion-cyclotron fluctuations, when $c_{s}<v_{A}$ (not shown). In contrast to three-wave resonance \cite{Sag&Gal69}, the revealed process can be named two-wave resonance. Similar two-wave resonance was previously reported between quasi-parallel whistler and electron-acoustic waves in the Earth's magnetosphere \cite{Vasko18:prl}. 

The electrostatic spikes observed in the simulations are basically identical to those observed aboard PSP. It is noteworthy that according to the simulations the spikes should occur only in the electrostatic component $E_{||}$ that is along the propagation direction. Since the observed low-frequency electromagnetic fluctuations propagate almost perpendicular to the $XY$ plane (Figure \ref{fig4}c), the spikes of about 2 mV/m observed aboard PSP (Figure \ref{fig2}) corresponds to a small projection of $E_{||}$ onto the $XY$ plane. The actual amplitude of the electrostatic spikes, whose major electric field is along the $Z$-axis and not measured by PSP, is expected to be one order of magnitude larger than observed in the $XY$ plane.

The parameters assumed in the Hall-MHD simulations ($\theta=5^{\circ}$, $\delta B/B_0=0.04$, and $f/f_{ci}\approx 1$) are quite typical of circularly-polarized low-frequency fluctuations observed in the solar wind \cite{Boardsen15:jgr,Zhao18:jgr,Liu23:apj_radial_evolution}, implying thereby that the revealed resonance can occur frequently. Several comments are in order about our simulations. First, the assumed polytrope indexes of $\gamma_{e}=1$ and $\gamma_{i}=3$ allowed reproducing the steepening of fast magnetosonic waves with $k\lambda_{i}\approx 0.75$ observed aboard PSP. Fully kinetic simulations would be needed to avoid the use of these {\it ad hoc} values. Note however that as long as $c_{s}>v_{A}$ the specific values of the polytrope indexes would not affect the steepening of fast magnetosonic waves with $k\approx k_{*}$. Second, we assumed a fixed amplitude of fast magnetosonic waves, while the steepening time scale $\tau_{S}\approx 40\;\omega_{ci}^{-1}$ is actually comparable with typical linear growth time $\tau_{\gamma}\approx 10^2$--10$^{4}\;\omega_{ci}^{-1}$ of such low-frequency fluctuations in the solar wind \cite{Verniero20,Klein21:apj}. Future simulations should incorporate the effect of linear growth. Third, wave breaking occurs in our simulations ($\nabla N/N\rightarrow \infty$ within finite time), because in the frame of Hall-MHD the ion-sound mode is basically dispersionless \cite{Formisano69}. The dispersion scale of the ion-sound mode is around electron Debye length $\lambda_{D}$ \cite{Sagdeev66}, whose typical value in the solar wind is $\lambda_{D}\approx 10^{-3}\lambda_{i}$. Before the dispersion set in, the steepening could, in principle, produce a thousand of harmonics in electric field spectra. In reality, dissipation processes set in much earlier, since the efficiency of Landau damping of ion-sound waves increases with frequency \cite{Ott&Sudan69,Dillard18:pop}.

We believe it is the dissipation processes (e.g., Landau damping by ions and electrons) that limit the number of harmonics in electric field spectra observed aboard PSP (Figure \ref{fig1}). The spatial scale of the electrostatic spikes remains much larger than local electron Debye length, because the dissipation processes can efficiently absorb the energy of high-frequency ion-sound fluctuations. This is a potential channel of energy transfer from cyclotron resonant ions driving circularly-polarized electromagnetic fluctuations to Landau resonant ions and electrons absorbing the energy of electrostatic fluctuations produced by the two-wave resonance. The presented results have wider applications, since electrostatic fluctuations associated with electromagnetic fluctuations, including whistler waves \cite{Agapitov18:grl,Vasko18:prl,An19:prl}, highly oblique fast magnetosonic waves \cite{Huang20:magnetosonic_harm}, electromagnetic ion-cyclotron waves \cite{Zhu&Chen19:emic_harmonics}, and kinetic
Alfv\'{e}n waves \cite{An21:jgr,Mozer&Vasko23:apj}, have been reported in other space plasma environments.

{\bf Acknowledgments:}
The of I.Y.V. was supported by the National Science Foundation grant No. 2026680. The work of X.A. and A.V. A. was supported by NASA grant No. 80NSSC22K1634. The work of F.S.M and I.V.K. was supported by NASA grant No. 80NSSC21K0730. J.V. acknowledges support from NASA LWS grant 80NSSC22K1014 and NASA PSP-GI grant 80NSSC23K0208. The data used in this study are publicly available at https://sprg.ssl.berkeley.edu/data/psp/.

\bibliographystyle{unsrt}
%\bibliography{full,full+}  %%% Remove comment to use the external .bib file (using bibtex).

\end{document}